%% file: amain-2Dmaterials-revised-nomarkup.tex
\documentclass[als,prl,twocolumn,groupedaddress,superscriptaddress,amsfonts,amssymb,amsmath,citeautoscript,a4paper]{revtex4-1}

\usepackage{lipsum}
\include{setup}

\usepackage{svg}
\usepackage{cleveref}
\usepackage{siunitx}

\usepackage{ulem}

\newcommand{\green}[1]{\textcolor{black}{#1}}

\DeclareSIUnit\wavenumber{\per\cm}

\usepackage[font=small,labelfont=bf,  justification=justified,format=plain]{caption}

\usepackage{gensymb}

\begin{document}

\title{Photon superbunching in cathodoluminescence of excitons in \ce{WS2} monolayer}

\author{Saskia Fiedler}
\thanks{S.~F. and S.~M. contributed equally to this work.}
\affiliation{POLIMA---Center for Polariton-driven Light--Matter Interactions, University of Southern Denmark, Campusvej 55, DK-5230 Odense M, Denmark}
\affiliation{Center for Nano Optics, University of Southern Denmark, Campusvej 55, DK-5230~Odense~M, Denmark}
\affiliation{AMOLF, Photonic Materials, Science Park 104, 1098 XG Amsterdam, The Netherlands}

\author{Sergii Morozov}
\thanks{S.~F. and S.~M. contributed equally to this work.}
\affiliation{POLIMA---Center for Polariton-driven Light--Matter Interactions, University of Southern Denmark, Campusvej 55, DK-5230 Odense M, Denmark}
\affiliation{Center for Nano Optics, University of Southern Denmark, Campusvej 55, DK-5230~Odense~M, Denmark}
\affiliation{CNG---Center for Nanostructured Graphene, Technical University of Denmark, DK-2800 Kongens Lyngby, Denmark}

\author{Leonid Iliushyn}
\affiliation{CNG---Center for Nanostructured Graphene, Technical University of Denmark, DK-2800 Kongens Lyngby, Denmark}
\affiliation{Department of Physics, Technical University of Denmark, DK-2800 Kongens Lyngby, Denmark}

\author{Sergejs Boroviks}
\affiliation{Center for Nano Optics, University of Southern Denmark, Campusvej 55, DK-5230~Odense~M, Denmark}
\affiliation{Nanophotonics and Metrology Laboratory, Swiss Federal Institute of Technology Lausanne (EPFL), Station 11, CH 1015, Lausanne, Switzerland}

\author{Martin Thomaschewski}
\affiliation{Center for Nano Optics, University of Southern Denmark, Campusvej 55, DK-5230~Odense~M, Denmark}
\affiliation{Department of Electrical \& Computer Engineering, The George Washington University, 800 22nd Street NW 5000 Science \& Engineering Hall, Washington, DC 20052, USA}

\author{Jianfang Wang}
\affiliation{Department of Physics, The Chinese University of Hong Kong, Shatin, Hong Kong SAR, China}

\author{Timothy J. Booth}
\affiliation{CNG---Center for Nanostructured Graphene, Technical University of Denmark, DK-2800 Kongens Lyngby, Denmark}
\affiliation{Department of Physics, Technical University of Denmark, DK-2800 Kongens Lyngby, Denmark}

\author{Nicolas Stenger}
\affiliation{CNG---Center for Nanostructured Graphene, Technical University of Denmark, DK-2800 Kongens Lyngby, Denmark}
\affiliation{Department of Electrical and Photonics Engineering, Technical University of Denmark, DK-2800 Kongens Lyngby, Denmark}

\author{Christian~Wolff}
\affiliation{POLIMA---Center for Polariton-driven Light--Matter Interactions, University of Southern Denmark, Campusvej 55, DK-5230 Odense M, Denmark}
\affiliation{Center for Nano Optics, University of Southern Denmark, Campusvej 55, DK-5230~Odense~M, Denmark}

\author{N.~Asger~Mortensen}
\email{asger@mailaps.org}
\affiliation{POLIMA---Center for Polariton-driven Light--Matter Interactions, University of Southern Denmark, Campusvej 55, DK-5230 Odense M, Denmark}
\affiliation{Center for Nano Optics, University of Southern Denmark, Campusvej 55, DK-5230~Odense~M, Denmark}
\affiliation{CNG---Center for Nanostructured Graphene, Technical University of Denmark, DK-2800 Kongens Lyngby, Denmark}
\affiliation{Danish Institute for Advanced Study, University of Southern Denmark, Campusvej 55, DK-5230~Odense~M, Denmark}

\begin{abstract}
\textbf{ABSTRACT} Cathodoluminescence spectroscopy in conjunction with second-order auto-correlation measurements of $g_2(\tau)$ allows to extensively study the synchronization of photon emitters in low-dimensional structures. Co-existing excitons in two-dimensional transition metal dichalcogenide monolayers provide a great source of identical  photon emitters which can be simultaneously excited by an electron. Here, we demonstrate large photon bunching with $g_2(0)$ up to $156\pm16$ of a tungsten disulfide monolayer (\ce{WS2}), exhibiting a strong dependence on the electron-beam current. To further improve the excitation synchronization and the electron-emitter interaction, we show exemplary that the careful selection of a simple and compact geometry -- a thin, monocrystalline gold nanodisk -- can be used to realize a record-high bunching $g_2(0)$ of up to $2152\pm236$. This approach to control the electron excitation of excitons in a \ce{WS2} monolayer allows for the synchronization of photon emitters in an ensemble, which is important to further advance light information and computing technologies. \\

\textbf{Keywords}: photon  superbunching, cathodoluminescence,  electron-beam excitation, transition metal dichalcogenide monolayer,  van der Waals heterostructures, hexagonal boron nitride
\end{abstract}

\maketitle

\section{Introduction}

The recent development in the synthesis and fabrication of two-dimensional (2D) transition metal dichalcogenide (TMDC) monolayers has sparked intensive research of their implementation in applications in the emerging field of quantum information and computing technologies~\cite{Aharonovich:2016,Reserbat-Plantey2021}. 
The family of TMDC monolayers with a thickness of a few atoms provides an attractive combination of optoelectronic properties, including a direct bandgap, a strong spin-orbit coupling and splitting, an optical access to spin and valley degrees of freedom~\cite{Manzeli2017}. 
An isolated monolayer is a robust material platform for  {photon} generation, where the co-existence of multiple excitons is a resource that represents an ensemble of identical photon sources.

Photon statistics of an ensemble of emitters localized within a sub-wavelength volume, strongly depends on the excitation synchronization and can exhibit photon bunching~\cite{Temnov:2009,Meuret:2015}. %
Photoexcitation does not provide the required synchronization of an ensemble as the photon arrival statistics is Poissonian in nature.
Moderate photon bunching -- with a $g_2(0)<2$ second-order auto-correlation function -- is characteristic for thermal light, for example the solar radiation of the Sun was measured to be $g_2(0)\simeq 1.6$~\cite{Tan2014}. 
The generation of superthermal light with $g_2(0)>2$ has been observed under the electron-beam excitation, namely in cathodoluminescence (CL)~\cite{Meuret:2015,Fiedler:2023}, which was theoretically described by the quantum master equation method~\cite{Yuge2022}.
 {The temporal correlation of CL photons with a Hanbury Brown and Twiss (HBT) interferometer gives an insight into coherent and incoherent processes in electron-matter interaction~\cite{Scheucher2022}}.
The width of the bunching peak provides information about the radiative lifetime of the emitters, while its amplitude indicates the probability of the excitation to interact with an ensemble of emitters~\cite{Meuret2020}.  

\begin{figure}[b]
	\includegraphics[width=1\linewidth]{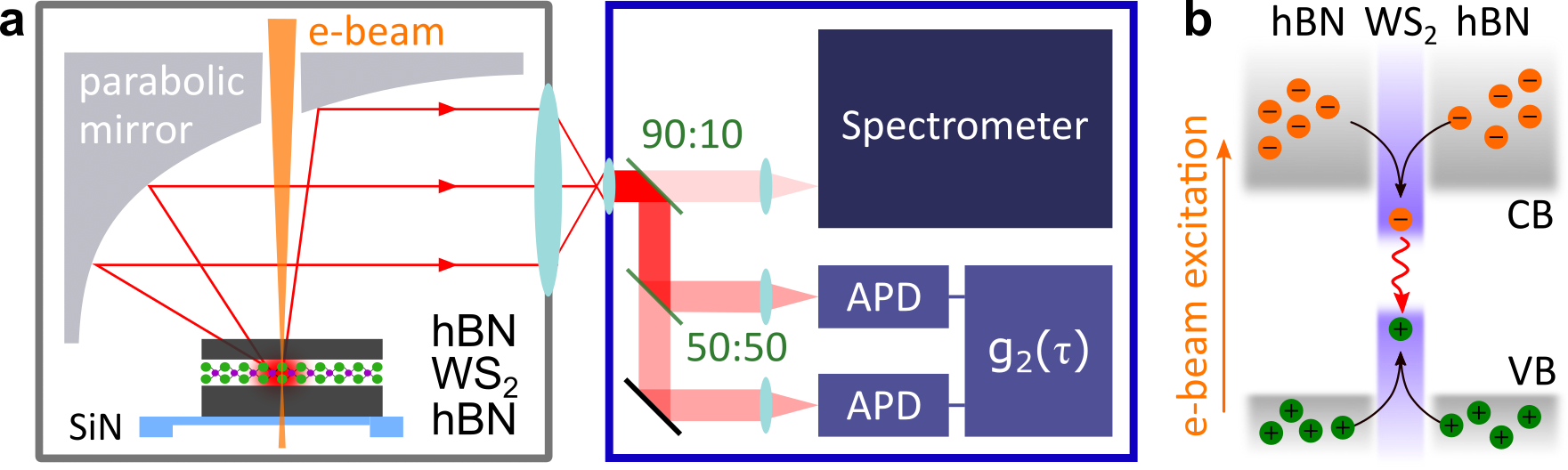}
	\caption{\textbf{Electron-beam excitation of \ce{hBN}-encapsulated \ce{WS2} monolayer.}
	\textbf{a}~Schematics of the experimental setup for detection of electron-beam-induced emission from a\ce{WS2} monolayer. The generated emission is collected with a parabolic mirror and directed towards a spectrometer and HBT interferometer consisting of two avalanche photodiodes (APD) and a 50:50 beamsplitter.
	\textbf{b}~Electron beam generates electron-hole pairs in \ce{hBN}, which relax to \ce{WS2} conduction and valence bands (CB and VB) with subsequent synchronized radiative recombination  {with spectral maximum around \SI{625}{\nano\meter} (1.984\,eV)}.
	}
	\label{fig1}
\end{figure}

In the limit of high electron-beam current, the electron arrival approaches Poissonian statistics, thus resulting in a flat second-order auto-correlation function $g_2(\tau)\to1$ as in the case of photoexcitation. 
However, reducing the electron-beam current to sub-few hundred~pA is allowing for a single electron excitation, which generates multiple electron-hole pairs capable of synchronized excitation of an ensemble. 
The synchronized radiative recombination does not necessarily imply phase locking, thus can be considered as simultaneous emission of emitters in an ensemble triggered by the same electron.
Exploiting the low currents, ensembles in various solid state systems – such as indium gallium nitride (\ce{InGaN}) quantum wells and color centers in nano-diamonds – have been shown to generate bunched light with $g_2(0)$ up to \num{80}~\cite{Meuret2018,Feldman:2018}.
A decrease of the electron-beam current to sub-pA is expected to further increase the $g_2(0)$-values, however such low currents are unattainable by common commercial scanning electron microscopes (SEMs).
Therefore, to further boost the photon bunching,  {ensembles consisting of identical and densely localized emitters within a sub-wavelength volume are needed \cite{Temnov:2009,Yuge2022}, while a geometry increasing the probability of electron-emitter interaction would increase the brightness of the source.}

In this letter, we study luminescence properties and photon statistics of a tungsten disulfide (\ce{WS2}) monolayer encapsulated in hexagonal boron nitride (\ce{hBN}) layers. 
{We report the first observation of photon bunching in a \ce{WS2} monolayer under electron-beam excitation, which reaches $g_2(0)=156\pm16$ at the lowest attainable current by our instrument. }
We further explore methods for improving the excitation synchronization and electron-emitter interaction via adding a monocrystalline gold nanodisk. 
In such a geometry,  {a gold disk can scatter and redirect incoming electrons to large angles, even allowing for excitation of in-plane momentum components. 
This electron excitation is increasing the probability of electron interaction with the sample, which allows us to} achieve record-high photon bunching of $g_2(0)=2152\pm236$, exceeding the previously reported values by over an order of magnitude.
Our simple geometry offers a robust and compact platform  {for efficient electron beam excitation and generation of light with photon superbunching statistics, which has important applications in light-based information and computing technologies.} 

\begin{figure}
	\includegraphics[width=1\columnwidth]{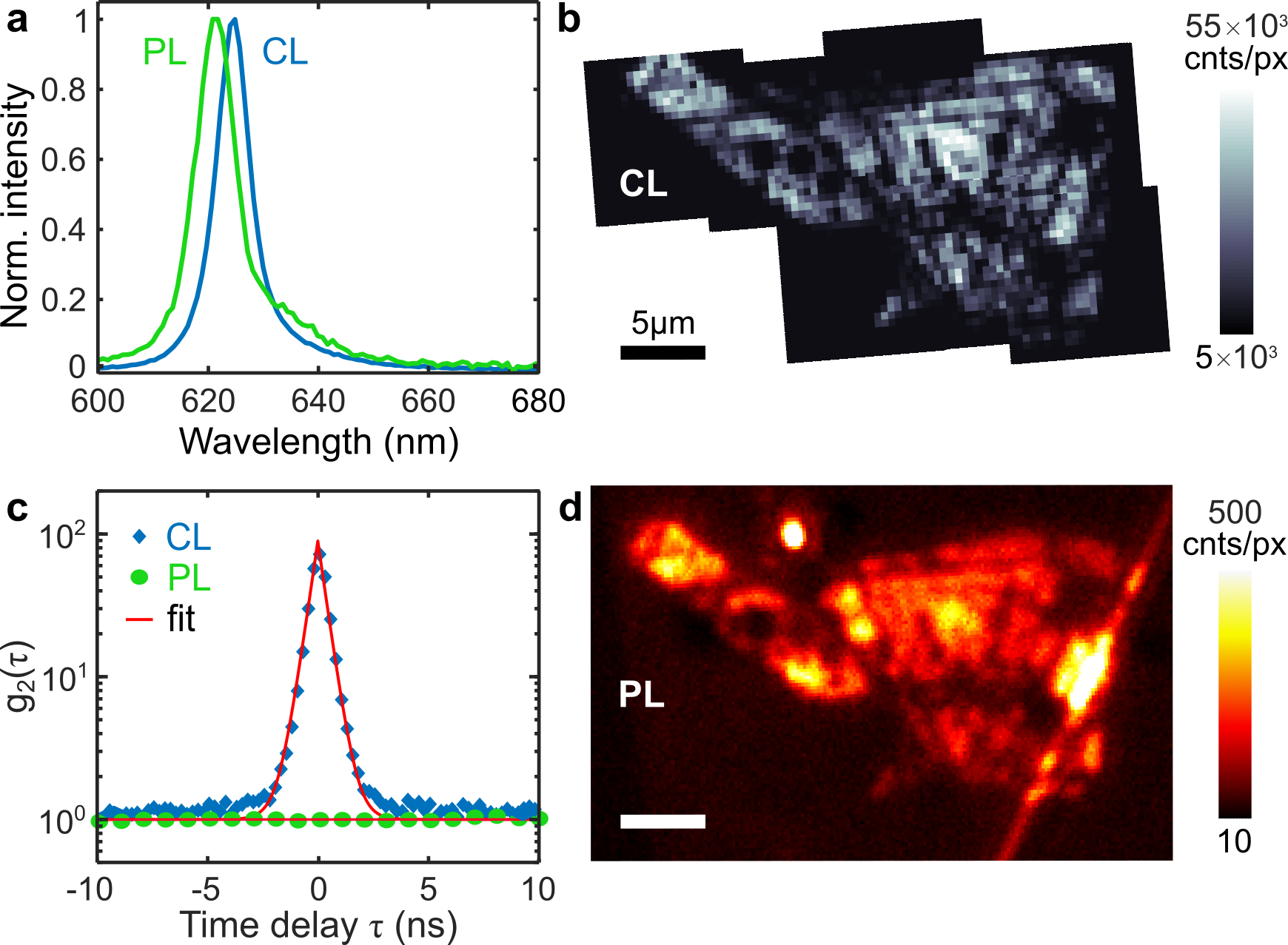}
	\caption{\textbf{Photo- and electron-beam excitation of hBN-encapsulated \ce{WS2} monolayer.}
	\textbf{a}~PL and CL spectra of \ce{WS2}--\ce{hBN} heterostructure.
	\textbf{b}~CL intensity map obtained by electron-beam scanning of the sample.	
	\textbf{c}~Electron-beam excitation synchronizes emitters, leading to the photon bunching, while photoexcitation only generates a flat $g_2(0)=1$.
	\textbf{d}~PL intensity map of the sample obtained by confocal scanning.
	}
	\label{fig2}
\end{figure}
\begin{figure*}
\includegraphics[width=0.9\textwidth]{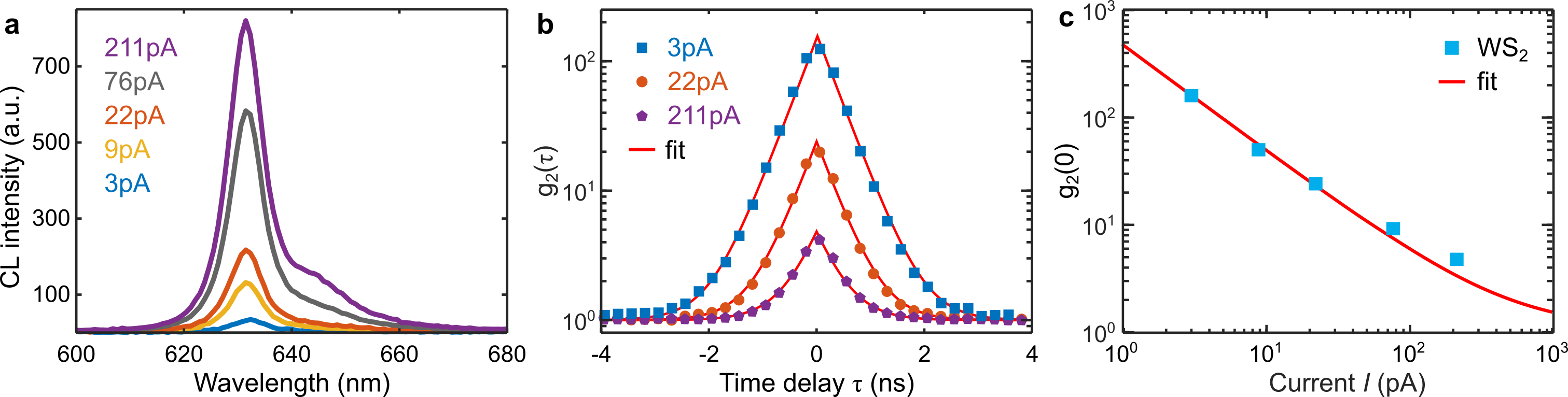}
	\caption{\textbf{Cathodoluminescence of hBN-encapsulated \ce{WS2} monolayer at increasing electron-beam current.}
	\textbf{a}~\SI{10}{\kilo\volt}-CL spectra  demonstrate the growth of CL intensity at  increasing electron-beam current. Red-shifted trion emission accompanies the neutral exciton at high electron-beam currents.
	\textbf{b}~Photon correlation histograms show reduction of photon bunching with increasing electron-beam current.
	\textbf{c}~Photon bunching factor $g_2(0)$ exhibits inverse dependence with electron-beam current.
	}
	\label{fig3}
\end{figure*}

\section{Results and discussion}

\subsection{Sample and experimental setup} 

We perform photoluminescence (PL) and cathodoluminescence (CL) experiments of a \ce{WS2} monolayer sandwiched between two thin \ce{hBN} flakes ({top and bottom flakes are ca. 35 and 105~nm thick, respectively, see also} SI Fig.~S1) {on a \SI{50}{\nano\meter}-thin silicon nitride (SiN) membrane}. 
The hBN encapsulation on both sides of the \ce{WS2} monolayer serves three purposes: (i) protecting the \ce{WS2} from electron-beam damage and contamination, (ii) increasing the effective interaction volume of incoming electrons with the monolayer, and (iii) providing layers for the generation of additional charge carriers which can subsequently diffuse into the \ce{WS2} and radiatively recombine, thereby significantly enhancing the CL intensity~\cite{Zheng:2017}.
{We note that even smallest contamination during the assembly of \ce{WS2}--\ce{hBN} heterostructure can cause quick and irreversible degradation of the \ce{WS2} monolayer under the electron beam.}
Fig.~\ref{fig1}a schematically presents the experimental setup for the electron-beam excitation of the investigated sample and for collection of generated CL. 
Here, the electron beam {of \SI{10}{\kilo\volt} and \SI{30}{\kilo\volt}} is focused onto the \ce{WS2}--\ce{hBN} heterostructure, and generated CL is subsequently collimated by a parabolic mirror and focused onto a spectrometer and an HBT interferometer (see details in SI). 

\subsection{Photo and electron-beam excitation}

Photo and electron-beam excitation are essentially different mechanisms of exciton generation in a \ce{WS2} monolayer.
An incoming photon typically creates one electron-hole pair in the \ce{WS2} monolayer, while an incoming electron can excite many electron-hole pairs in the sample (Fig.~\ref{fig1}b), predominantly in the thicker \ce{hBN} layers due to a large interaction volume~\cite{Yacobi:1990}. 
The latter process can be described by a scattered incoming electron which excites high-energy interband transitions in the sample, also referred to as bulk plasmons in electron spectroscopy literature~\cite{Egerton:2011,Meuret:2015}.
Such bulk plasmons~{($\sim30$~eV)} are comprised of a coherent sum of interband transitions contrary to the collective intraband oscillations of free charges, e.g., manifesting in surface plasmons~{($< 5$~eV)}.
Electron induced bulk plasmons subsequently decay into multiple  electron-hole pairs which, after a typical excitonic lifetime, exhibit synchronized radiative recombination. 
The photons appear to arrive as a "packet" and can be detected with the HBT interferometer as photon bunching ($g_2(0) > 1$).

In Fig.~\ref{fig2}a, we compare spectral emission properties of the \ce{WS2}--\ce{hBN} heterostructure under photo and electron-beam excitation.
The PL spectrum is centered around \SI{621}{\nano\meter} and has a red-shifted satellite peak at \SI{635}{\nano\meter} (green line in  Fig.~\ref{fig2}a). 
These emission lines are characteristic for the neutral and red-shifted charged exciton (trion) in \ce{WS2} monolayers at room temperature \cite{Morozov2021}. 
The CL spectrum in Fig.~\ref{fig2}a is dominated by the neutral exciton peak at \SI{626}{\nano\meter} which is red-shifted with respect to the PL spectrum. 
In general, we observe a sub-\SI{10}{\nano\meter} spectral wandering of the excitonic peak in both PL and CL measurements which is possibly due to local strain within the  {\ce{WS2}--\ce{hBN} heterostructure}~\cite{Li2020}.
Fig.~\ref{fig2}b presents a CL intensity map obtained by scanning the electron beam over the \ce{WS2}--\ce{hBN} heterostructure. The spatial CL distribution reveals areas of lower intensity, which are due to cracks in the \ce{WS2} monolayer. %
The round dark patches apparent in the CL map can be attributed to bubbles in between the three layers, as the lack of adherence in the stacked sample has been shown to result in quenched luminescence~\cite{Nayak:2019}.
Fig.~\ref{fig2}d displays a corresponding PL confocal map obtained with laser excitation at \SI{404}{\nano\meter}. We find a similar intensity distribution as in the CL map in Fig.~\ref{fig2}b, although characterized by a lower spatial resolution than with an electron beam.

The photon statistics of emission generated from the \ce{WS2}--\ce{hBN} heterostructure drastically depend on the type of excitation.
Fig.~\ref{fig2}c presents photon correlation histograms obtained under photo and electron-beam excitation. 
Electron-beam excitation results in a pronounced photon bunching peak at zero correlation time (blue diamonds in Fig.~\ref{fig2}c), while the photon correlation histogram obtained under photoexcitation is flat with $g_2(\tau) = 1$ (green circles in Fig.~\ref{fig2}c). 
Moreover, the bunching amplitude in CL can be tuned by the electron-beam current, unlike the flat photon correlation histogram in PL \cite{Meuret:2015}.

\subsection{CL response to electron-beam current}

The CL intensity and the amplitude of photon bunching can be controlled by the electron-beam current $I$. 
As the current is increased, a higher CL intensity is observed due to the larger amount of incoming electrons, generating more electron-hole pairs in the sample, which can subsequently recombine radiatively in the \ce{WS2} monolayer (Fig.~\ref{fig3}a). 
In contrast, the amplitude of the photon bunching peak follows the opposite trend: with increasing $I$, the contribution of uncorrelated emission is increased, resulting in a lower photon bunching \green{(SI Fig.~S2)}. 
Fig.~\ref{fig3}b shows the bunching histograms collected at low and high electron-beam currents, which we fit to extract the values of $g_2(0)$ (see details in SI). 
At the lowest electron-beam current of \SI{3}{\pico\ampere}, we achieve a giant bunching factor $156\pm16$, indicating the high quality of the investigated sample. 
In contrast, the bunching factor is greatly reduced to approximately $4.7\pm0.2$ for the highest electron-beam current of \SI{211}{\pico\ampere}, which is summarized in Fig.~\ref{fig3}c.
We fit the examined dynamics with 
\begin{equation}
g_2(0,I)=1-\frac{I_0}{P_{e} \times I}
\label{eq}
\end{equation}
as the bunching factor has been observed to be inversely dependent on the applied electron-beam current~\cite{Meuret:2015, Meuret:2020}. 
Here $I_0$ relates to the current exciting a bulk plasmon and $P_e$ is a probability of incoming electrons to interact with emitters.

{The physical reasons behind the photon bunching can be attributed to the synchronization of emission from emitters, where each incoming electron creates a photon packet.
With increasing electron-beam current, the electrons arrive closer in time and the photon packets become increasingly indistinguishable, thereby reducing the bunching factor until the Poissonian distribution $g_2(0,{I\rightarrow}\infty) = 1$ is reached.}
In the \ce{WS2}--\ce{hBN} heterostructure, the inverse $I$-dependence is confirmed in Fig.~\ref{fig3}c,  although the Poissonian distribution cannot be reached due to sample degradation at high electron-beam currents $I >$~\SI{200}{\pico\ampere} \green{(SI Fig.~S3)}. 

\begin{figure*}
	\includegraphics[width=0.9\textwidth]{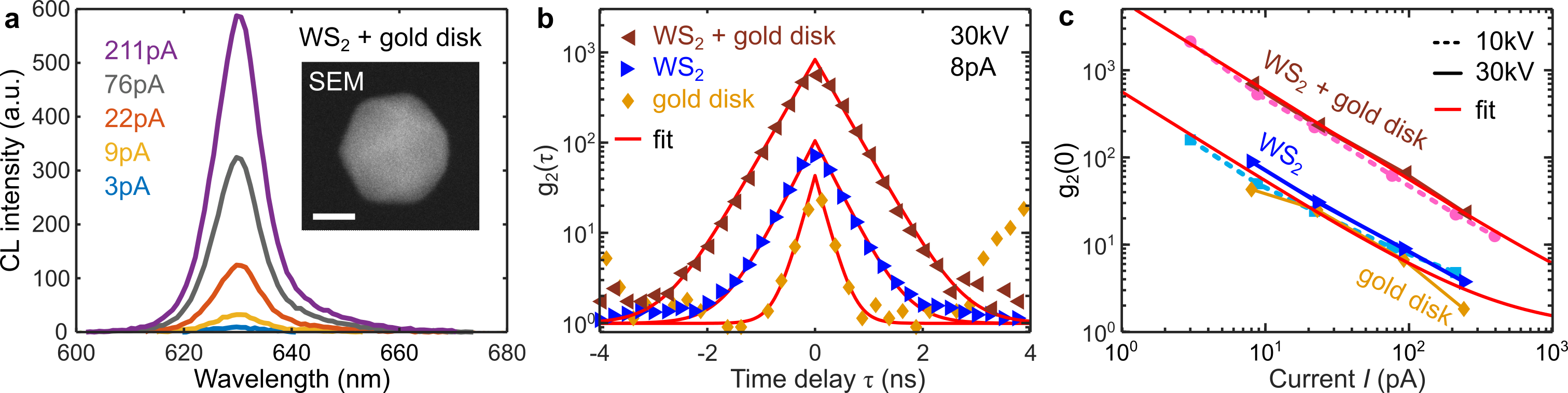}
	\caption{\textbf{Enhancing electron beam interaction with the sample using a gold nanodisk. }
	\textbf{a}~\SI{10}{\kilo\volt}-CL spectra of \ce{hBN}-encapsulated \ce{WS2} monolayer excited through a gold nanodisk with a diameter of \SI{120}{\nano\meter} at increasing electron-beam current. The inset shows an SEM image of the gold nanodisk, the scale bar denotes \SI{50}{\nano\meter}.
	\textbf{b}~Photon correlation histograms of \ce{hBN}-encapsulated \ce{WS2} monolayer (blue triangles), \ce{hBN}-encapsulated \ce{WS2} monolayer with a gold nanodisk on top (dark red triangles), and a similar \SI{120}{\nano\meter}-sized gold nanodisk on \ce{hBN} (orange diamonds).  {The long tails at 4~ns is a detection artifact caused by APD after-glow \green{(see SI Fig.~S6)}.}
	\textbf{c}~Photon bunching factor $g_2(0)$ vs. electron-beam current $I$ at \SI{10}{\kilo\volt} (dashed lines) and \SI{30}{\kilo\volt} (solid lines) for the three samples as in \textbf{b}.
	}
	\label{fig4}
\end{figure*}

\subsection{Boosting bunching through geometry}

A further increase of the photon bunching factor by reducing the applied electron-beam current is limited by the instrument. 
Decreasing the acceleration voltage potentially allows to reduce the current to sub-\si{\pico\ampere}, however in practice, it causes severe sample charging already at \SI{5}{\kilo\volt}. 
To overcome this limit, we suggest to modify the geometry by adding a thin gold nanodisk to induce a local change of electron-beam excitation parameters.
We drop-cast monocrystalline gold nanodisks with a diameter ranging from \SI{80}{\nano\meter} to \SI{230}{\nano\meter} on top of the \ce{WS2}--\ce{hBN} heterostructure. 
{Gold nanodisks are well separated from each other across the heterostructure, allowing for a comparison of electron-beam excitation of bare heterostructure and through a gold nanodisk {(SI Fig.~S1)}.}

For further CL studies, we chose a gold nanodisk with an approximate  {thickness of \SI{30}{\nano\meter} and a} diameter of \SI{120}{\nano\meter}, as shown in the inset of Fig.~\ref{fig4}a.
The reasoning behind this selection is that the (broad) plasmonic dipole mode spectrally overlaps with the \ce{WS2}-exciton, in principle allowing for plasmonic interaction \cite{Fiedler:2020}.
Fig.~\ref{fig4}a presents the CL spectra of the \ce{WS2}--\ce{hBN} heterostructure excited through the gold nanodisk with increasing electron-beam current. We observe an overall CL brightening with increasing $I$, similar to the bare \ce{WS2}--\ce{hBN} heterostructure. However, the CL spectra do not reveal  {significant} emission enhancement which would indicate a plasmon-assisted Purcell effect, nor any evidence of strong coupling (Fig.~\ref{fig4}a). This observation is in agreement with the PL measurements, where the position of the gold nanodisks distributed over the entire \ce{WS2}--\ce{hBN} heterostructure cannot be made out.
We ascribe this lack of luminescence enhancement to the large separation of about \SI{35}{\nano\meter} between the gold nanodisk and the \ce{WS2} monolayer, which does not allow for efficient near-field coupling. 

Remarkably, the addition of gold nanodisks improves the emission synchronization in the \ce{WS2}, which manifests in the increase of photon bunching factor. 
The gold nanodisks themselves -- on \ce{hBN} only -- exhibit dim CL emission \green{(see SI Fig.~S3-S4)}, which allowed us to quantify the bunching factor $g_2(0)=43.1\pm 4.9$ of a similarly sized \SI{120}{\nano\meter} nanodisk (orange in Fig.~\ref{fig4}b). Fig.~\ref{fig4}b further compares the photon correlation histograms measured at identical excitation parameters (\SI{8}{\pico\ampere}, \SI{30}{\kilo\volt}) from the bare \ce{WS2}--\ce{hBN} heterostructure (blue), and a system of a gold nanodisk on  \ce{WS2}--\ce{hBN} heterostructure (dark red). 
The fit of the histograms provides information about the lifetime of the generated emission $\tau$ and the bunching factor $g_2(0)$. 
We observe a minor change in radiative lifetime of \ce{WS2} after the addition of the gold nanodisk, extracting values of  $\tau_{\ce{WS2}} = 365\pm18$~\si{\pico\second} and $\tau_{\ce{WS2}+\ce{gold}} = 416\pm10$~\si{\pico\second}.
The extracted values of radiative lifetime agree with one obtained via time-correlated single-photon counting (TCSPC) methods using a pulsed laser ($\tau^{\rm PL}_{\ce{WS2}} = 296\pm22$~\si{\pico\second}, \green{see SI Fig.~S5}). 
We note that the reported lifetimes here do not represent the intrinsic exciton decay rate in \ce{WS2}, and are effective radiative lifetimes depending on many factors such as optical environment, available decay channels etc.~\cite{Wang2018}.    
We conclude that the addition of a gold nanodisk did not facilitate the creation of additional, faster relaxation channels for the \ce{WS2}-excitons which is in agreement with the absence of Purcell enhancement. For the bare gold nanodisk, however, we measure a lifetime of $\tau_{\ce{gold}} = 196\pm34$~\si{\pico\second}, which is close to the temporal limit of our setup (\SI{180}{\pico\second}). 

We extract a photon bunching factor of $105\pm11$ for the bare \ce{WS2}--\ce{hBN} heterostructure, which greatly increases to $829\pm114$ for the system of a gold nanodisk on the \ce{WS2}--\ce{hBN} heterostructure at 30~kV. A simple addition of the bunching factors of the bare gold nanodisk and that of the \ce{hBN}-encapsulated \ce{WS2} cannot explain the huge increase of bunching to $829\pm114$ (which we will focus on in the next section). The highest bunching factor of $2152\pm236$, we observe at the lowest electron-beam current of \SI{3}{\pico\ampere} (at \SI{10}{\kilo\volt}) for the gold nanodisk on the \ce{WS2}--\ce{hBN} heterostructure (pink circles in Fig.~\ref{fig3}c).
Finally, we confirm the inverse electron-beam current dependence of the photon bunching using a different geometry and a reduced acceleration voltage (from \SI{30}{\kilo\volt} to \SI{10}{\kilo\volt} allowing for a further reduction in $I$). We plot the summary of the results in Fig.~\ref{fig4}c, including fits of the $g_2(0,I)$-data with an inverse current function (red curves).  {We extract from the fit a large change in $I_0$ from $I_0^{\ce{WS2}}/P_e=477\pm14$\,pA to $I_0^{\ce{WS2}+\ce{gold}} /P_e =5257\pm154$\,pA, which indicates a drastic change in the probability of electrons to interact with the sample due to the presence of a gold nanodisk.}

\begin{figure*}
	\includegraphics[width=0.99\linewidth]{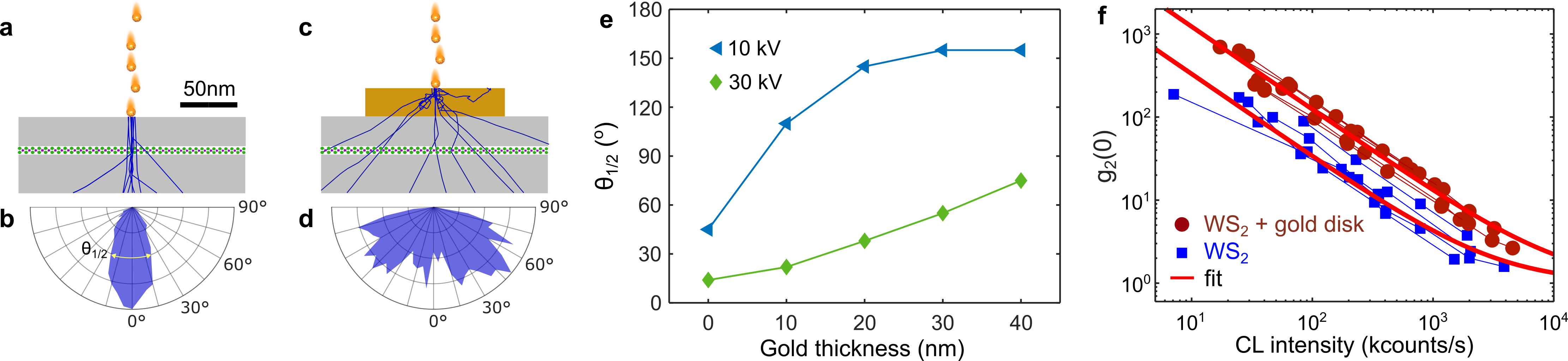}
	\caption{
	\textbf{Diverging electrons.}
	 {\textbf{a,c} Monte Carlo simulations of electron trajectories at \SI{10}{\kilo\volt} in a bare \ce{hBN}--\ce{WS2} sandwich and with a gold disk on top. 10 trajectories (blue traces) are shown for each case. 
	\textbf{b,d}~Angular diagrams of electron divergence $\Theta_{1/2}$ for the geometries presented in panels \textbf{a,c}.
	\textbf{e}~$\Theta_{1/2}$ increases with the thickness of gold nanodisk as simulated at 10 and 30~kV acceleration. 
	\textbf{f}~$g_2(0)$-intensity dependencies for 5 gold nanodisks as in \textbf{c} (dark red circles) and  reference points of bare \ce{WS2}--\ce{hBN} heterostructures as in \textbf{a} (blue squares) measured within 10~$\mu$m from the corresponding gold nanodisk. The higher CL emission intensity in the case of through-gold excitation indicates a higher probability of bulk plasmon excitation due to the presence of gold nanodisks. 
	}
	}
	\label{fig5}
\end{figure*}

\subsection{ {The role of gold nanodisk}}

We perform Monte Carlo simulations~\cite{Drouin:2007} to shed light on  {the change in electron interaction probability and photon superbunching} caused by the addition of a gold nanodisk.
Fig.~\ref{fig5}a,c present results for electron trajectory simulations of the investigated geometries using a \SI{10}{\kilo\volt}-electron beam with a spot size of \SI{5}{\nano\meter}. 
In case of the bare \ce{WS2}--\ce{hBN}  heterostructure, the hBN interface barely affects the direction of incoming electrons, which is visualized using 10 simulated electron trajectories in Fig.~\ref{fig5}a (blue lines). We further accumulate 1000 trajectories to calculate the electron angular divergence histogram (Fig.~\ref{fig5}b). Evidently, it reveals the principal direction of electron propagation normal to the sample plane with a small divergence of $\Theta_{1/2}^{\ce{WS2}} = 45 \degree$ (the full width at half power). 
The addition of a 30~nm thick gold nanodisk in Fig.~\ref{fig5}c drastically changes the direction of electron propagation within the \ce{WS2}--\ce{hBN}  heterostructure as it is shown in the corresponding angular divergence histogram in Fig.~\ref{fig5}d. 
The incoming electrons are widely scattered by the gold nanodisk so the principal direction of electron propagation cannot be identified, while electrons propagate omnidirectionally with divergence $\Theta_{1/2}^{\ce{WS2}+\ce{gold}} = 155 \degree$. 
We summarize the simulation results in Fig.~\ref{fig5}e, which presents the electron divergence for a range of gold thickness and acceleration voltage.

The divergence of the electron beam serves to (on average) provide impinging electrons with a significant in-plane momentum which further enhances coupling to the in-plane dipole moment of the excitons.
Moreover, the redirection of electrons by a gold nonodisk from the normal-to-sample plane extends the geometrical path of the electrons in the \ce{WS2}--\ce{hBN} heterostructure. This increases the probability of electrons $P_e$ to interact with the thin sub-100~nm sample and to excite bulk plasmons in it. %
As the interaction length and the probability of bulk plasmon excitation are both increased, one would expect to observe an increase in CL emission intensity while preserving the degree of bunching in the experiment. 
Fig.~\ref{fig5}{f} presents the results of such an experiment, where we collect both the $g_2(0)$ and the corresponding CL intensity from the bare heterostructure and that with an additional gold nanodisk.
We average the $g_2(0)$-intensity data sets in Fig.~\ref{fig5}{f} and fit the result with an intensity inverse function since the CL intensity is proportional to the electron-beam current. The fit results clearly demonstrate the brightening of CL intensity after adding a gold nanodisk, which   is due to the increase in probability of electrons to interact with the sample. Such a change in excitation efficiency improves the synchronization of an ensemble, which results in increase of the degree of photon bunching according to Eq.~\ref{eq}~\cite{Meuret:2015,Yuge2022}, which we observe in Fig.~\ref{fig4}c.

\section{Conclusion and Outlook}

In conclusion, we demonstrated the importance of electron-beam excitation for synchronization of large emitter ensembles, and for achieving high photon bunching factors. We reported  {the first experimental observation of photon bunching  $g_2(0)=156\pm16$ in cathodoluminescence of excitons in \ce{WS2} monolayer}. 
We further suggested that an appropriate geometry can locally  {facilitate} the electron-beam interaction with the sample, which  {is allowing to increase the CL intensity signal as well as} to boost the photon bunching dynamics far beyond the instrumentation-limited currents. We utilized a system of a thin gold nanodisk on a \ce{WS2}--\ce{hBN} heterostructure to demonstrate  {a dramatic increase in photon bunching to} $g_2(0)=2152\pm236$, which is over an order of magnitude larger than previously reported values.
Employing Monte Carlo simulations in combination with CL-$g_2(\tau)$-measurements, we showed  {that a gold nanodisk can redirect incoming electrons in the sample increasing the interaction length and affecting the bulk plasmon excitation probability, which is another parameter allowing to manipulate the photon bunching dynamics beside the electron-beam current.}
We presented a simple way to drive the values of $g_2(0)$ to extremes, which offers a  {photon source with super-Poissonian emission statistics for applications in emerging light information and computing technologies}. 

\section{Supporting Information}
{The Supporting Information includes details on sample fabrication; 
cathodoluminescence spectroscopy;
photon correlation measurements;
imaging of \ce{hBN}-encapsulated \ce{WS2} monolayer; 
uncorrelated emission contribution at high electron-beam currents; 
analysis of CL intensity at low and high electron-beam currents;
CL spectrum of a gold nanodisk;
lifetime measurements;   
Raman characterization of the sample.}

\section{Acknowledgments} 

The authors gratefully acknowledge the experimental support of V.~Zenin. 
N.~A.~M. is a VILLUM Investigator supported by VILLUM FONDEN (Grant No.~16498).
S.~M. acknowledges funding from the Marie Sk\l{}odowska-Curie Action (Grant agreement No.~101032967).
C.~W. acknowledges funding from a MULTIPLY fellowship under the Marie Sk\l{}odowska-Curie COFUND Action (Grant agreement No.~713694).
The Center for Nanostructured Graphene (CNG) is sponsored by the Danish
National Research Foundation (Project No.~DNRF103).
The Center for Polariton-driven Light--Matter Interactions (POLIMA) is sponsored by the Danish National Research Foundation (Project No.~DNRF165).

\section{Author contributions}

S.~F. conceived the idea and managed the project. 
S.~M. and S.~F. designed and built the experimental setup.
S.~F., S.~M., and M.~T. performed optical measurements; S.~B. and S.~F. conducted Raman characterization. 
L.~I. assembled heterostructures; J.~W. synthesized monocrystalline gold nanodisks. 
S.~F. performed the Monte Carlo simulations of electron trajectories. 
S.~B. and S.~M. prepared the cover art.
S.~M., S.~F., and C.~W. analyzed the data. 
The project was supervised by N.~A.~M., T.~J.~B., C.~W., and N.~S.
The results were discussed by all authors and the writing of the manuscript was done in a joint effort. 
All authors provided critical feedback and helped shape the research, analysis, and manuscript.

\bibliographystyle{unsrt}

\bibliography{references}

\newpage
\begin{figure*}
  \centering
  \includegraphics[width=8.25cm]{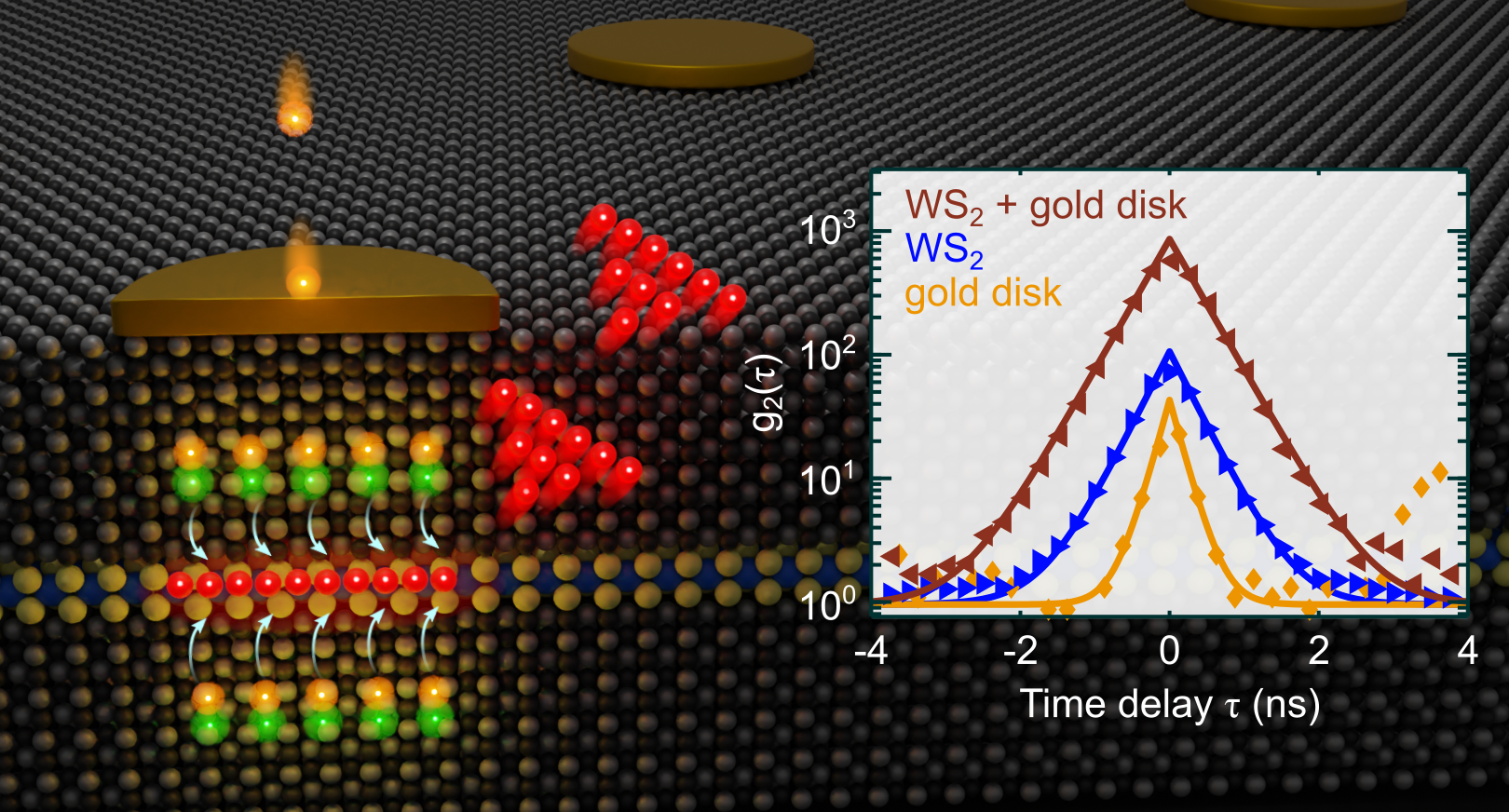}
  \caption*{\textbf{TOC Graphic.} Artistic view for generation of highly bunched light in cathodoluminescence. A high-energy electron generates multiple electron-hole pairs in \ce{hBN} layers, which experience synchronized radiative recombination in the sandwiched \ce{WS2} monolayer.}
\end{figure*}

\end{document}

%% file: setup.tex
\usepackage[utf8]{inputenc}
\usepackage[english]{babel}
\usepackage{microtype}
\usepackage{txfonts}
\usepackage{txfontsb}
\usepackage{bm}
\usepackage{xcolor}
\usepackage{graphicx}
\usepackage{float}
\usepackage{xspace}

\usepackage{enumerate}
\usepackage[inline]{enumitem}

\usepackage[version=4]{mhchem}

\usepackage{siunitx}
\DeclareSIUnit[number-unit-product=]\percent{\char`\%} 

\DeclareSIUnit\wavenumber{1/cm⟩}

\usepackage[pdftex]{hyperref}
\hypersetup{colorlinks,
            linkcolor={blue!75!black!80!yellow},
            citecolor={blue!75!black!80!yellow},
            urlcolor={blue!75!black!80!yellow},
            pdfstartview=FitH}
            
\usepackage{xifthen}
\usepackage{etoolbox}
\usepackage{soul}

\usepackage{mdframed}
\definecolor{blue-violet}{rgb}{0.54, 0.17, 0.89}

\newmdenv[topline=false, rightline=false, bottomline=false,%
  linewidth=1.25pt, innerrightmargin=0pt, leftmargin=-8pt,%
  innerleftmargin=7pt, skipabove=0pt, skipbelow=0pt,%
  linecolor=blue-violet, fontcolor=blue-violet]{mdleftbar}